
\raggedbottom


\def\nex{\par\noindent\hang}

\null
\vskip 20mm
\centerline{\bf  Inverse Dynamical Population Synthesis and Star Formation}

\vskip 2mm

\centerline{\bf K1}

\vskip 10mm
\centerline{\bf Pavel Kroupa}
\vskip 10mm
\centerline{Astronomisches Rechen-Institut}
\vskip 2mm
\centerline{M{\"o}nchhofstra{\ss}e~12-14, D-69120~Heidelberg, Germany}
\vskip 10mm
\centerline{e-mail: S48@ix.urz.uni-heidelberg.de}

\vfill

\centerline{MNRAS, in press}

\vfill\eject

\hang{
\bf Abstract.
\rm
Recent observations of pre-main sequence stars suggest that all stars may form
in multiple systems. However, in the Galactic field only about 50~per~cent of
all systems are binary stars. We investigate the hypothesis that stars form in
aggregates of binary systems and that the dynamical evolution of these
aggregates leads to the observed properties of binary stars in the Galactic
field. A thorough analysis of star count data implies
that the initial stellar mass function rises
monotonically with decreasing mass and that it can be approximated by three
power law segments. Together with
our assumption that the birth mass-ratio distribution is not correlated this
leads to a contradiction with the distribution of
secondary masses in Galactic field binaries with G-dwarf primaries which have
too few low-mass companions. For the inverse dynamical
population synthesis we assume that the initial distribution of periods is
flat in log$_{10}P$, where $P$ is the orbital period in days, and
$3\le $log$_{10}P\le 7.5$. This is consistent with pre-main sequence data.
We distribute $N_{\rm bin}=200$~binaries in aggregates
with half mass radii $0.077\le R_{0.5}\le 2.53\,$pc, corresponding to the
range from tightly clustered to isolated star formation, and follow the
subsequent evolution of the stellar systems by direct N-body integration.
We find that
hardening and softening of binary systems do not significantly increase the
number of orbits with log$_{10}P<3$ and log$_{10}P>7.5$, respectively. After
the cluster with $R_{0.5}\approx0.8\,$pc disintegrates we
obtain a population which consists of about 60~per~cent binary systems
with a period distribution for log$_{10}P>4$ as is observed and in which the
G-dwarf binaries have a mass ratio distribution which agrees with the observed
distribution. This result indicates that the majority of
Galactic field stars may originate from a clustered star formation mode,
characterised by the {\it dominant mode cluster} which has initially $(N_{\rm
bin},R_{0.5})\approx(200,0.8\,{\rm pc})$. We invert the orbit depletion
function and
obtain an approximation to the initial binary star period distribution for star
formation in the dominant mode cluster. Comparison with the measured
distribution of orbits for pre-main sequence stars formed in the distributed
mode of star formation suggests that
the initial distribution of binary star orbits may
not depend on the star formation environment.
If a different stellar mass function to the one we adopt is assumed then
inverse dynamical population synthesis cannot solve for an aggregate in
which the initial binary star population
evolves to the observed population in the Galactic field. This implies
that the Galactic field stellar mass function may be related
to the stellar density at birth in the most common, or dominant, mode of star
formation.

\vskip 5mm

{\bf Keywords:} stars: low mass, mass function  -- binary stars: orbital
parameters, evolution -- stellar clusters: Galactic, embedded --
star formation: distributed, clustered, dominant mode
}

\vfill\eject

\noindent{\bf 1 INTRODUCTION}
\vskip 12pt
\noindent
Although the formation
of a single star appears to be understood fairly well a stimulating debate
continues as to the formation mechanism of multiple star systems
(see e.g. Bodenheimer et
al. 1993). This has become a very important issue because during the recent
two years observations
have shown that the orbital period distribution and frequency of occurrence
of pre-main sequence binary stars differs from that observed on the main
sequence suggesting that virtually all pre-main sequence stars may be in binary
systems.

Abt \& Levy (1976, see also Abt 1987) pioneered the detailed study of
statistical
properties of orbital parameters of binary systems with a solar-type primary.
Many of their findings are verified and improved by the
long-term CORAVEL spectroscopic survey of nearby solar-type
stars published by Duquennoy \& Mayor (1991).
Combined with data on visual binaries and common proper motion
systems this survey has established the following
four points of particular interest for the present study: 1)~the proportion,
$f_{\rm G}$, of binary systems with a G dwarf primary is about
50--60~per~cent,
2)~the distribution of orbital periods, $P$, is approximated rather well by a
Gaussian distribution in log$_{\rm 10}P$
(in what follows we measure $P$ in days)
with mean $\overline{{\rm log_{10}}P}=4.8\,$, 3)~the
eccentricity distribution for systems with orbital period log$_{\rm
10}P>3$ approximately follows the thermal distribution but is bell shaped for
smaller periods, and 4)~the mass ratio
distribution does not follow the expected distribution if the secondaries were
distributed
according to the field-star mass function: low-mass secondaries
are underrepresented (Kroupa \& Tout 1992). Systems with periods
log$_{10}P>3$ have a mass ratio distribution that increases with decreasing
mass ratio but flattens with a possible decline below a mass ratio of
approximately~0.25. For smaller periods the mass ratio distribution is flat or
might even decrease with decreasing mass ratio (Mazeh et al. 1992).
Similar results appear to hold for K-dwarf stars (Mayor et al. 1992) and
M~dwarfs (Fischer \& Marcy 1992).

Occultation observations of pre-main sequence stars in the Taurus star-forming
region supplemented by infrared imaging show that the proportion of binaries
among pre-main sequence stars with separations between about 2~AU and 1400~AU
is approximately 1.5~times larger than on the main sequence (Simon 1992). A CCD
imaging survey of visual pre-main sequence binaries in southern dark
clouds, all at a distance of about $150\,$pc, also suggests that the proportion
of binaries among pre-main sequence stars is about 1.5~times larger
than on the main
sequence in the apparent separation range of 150 to 1800~AU
(Reipurth \& Zinnecker 1993). An infrared speckle
interferometry survey of young stars in the Taurus star forming region,
combined with direct imaging and lunar occultation observations, shows that
approximately two times as many binaries are found in the apparent distance
range of 18~to 1800~AU
in the Taurus star-forming region than on the main
sequence (Leinert et al. 1993 and Richichi et al. 1994). A similar study
of the Taurus-Auriga and Ophiuchus-Scorpius star-forming regions
by
Ghez, Neugebauer \& Matthews (1993), on the other hand, indicates a factor of
four times as many pre-main sequence binaries
than main sequence stars in the apparent separation range 16 to 252~AU.
Their incompleteness corrections differ from those applied by
the other groups, in that they attempt to account for missed companions fainter
than the primary star by more than two K-band magnitudes (for a further
discussion of the incompleteness corrections and pre-main sequence binary
stars see Mathieu 1994).
Mathieu (1992) reports on radial velocity measurements of pre-main sequence
stars and finds that the proportion of these that are binaries with
$P<100\,$days
(semi-major axis less than 0.5~AU if the mass of the system is
$1.4\,M_\odot$) is indistinguishable from the proportion on the main sequence.

The above studies suggest that the distribution of orbital elements
for very young systems differs from those on the main sequence. It would
appear
that the form of the period distribution evolves as well as the number of
binaries.
The two mechanisms by which orbital parameters of binary systems evolve
are:
1) Interaction of the two proto stars with each other and with the proto
stellar material that remains in the system ({\it pre-main sequence
eigenevolution}), and 2) perturbations
caused by close passages of other systems ({\it stimulated evolution}). The
first
mechanism becomes increasingly important for systems with smaller separation of
components and will be most effective in systems that are younger than the
evolution time of the circumproto-stellar material which is roughly
$10^5$~yrs. This mechanism implies
a more or less continuous evolution of orbital parameters. The
second mechanism, on the other hand, is effective on a time scale characterised
by the rate of evolution of the cluster of young stars and is increasingly
effective for systems with larger separation of components. The close passage
of another system leads to abrupt changes of the orbital elements. In this
paper we are concerned with stimulated evolution only.

Past simulations of interactions
of binaries with single stars show that these can lead to ionisation of
the binary, or to changes in orbital elements and exchange of stellar
components. In particular, Heggie (1975) developed in detail the theoretical
understanding of the reaction of binary stars to perturbations by other
systems and highlights the complexity of the interactions of higher
order systems, and Hills
(1975) performed extensive numerical simulations of such
processes. Both studies show that tightly bound binary stars tend to heat the
field by gaining binding energy, whereas relatively wide systems are ionised
rapidly without affecting the energy content of the surrounding field
significantly.

In this paper we
study the evolution of the distribution function of periods,
eccentricities and mass ratios of a population of binary stars that
are initially grouped in an aggregate. We adopt initial distributions that are
consistent with all available pre-main sequence observational constraints. By
fitting the final model distribution functions to observed Galactic field
distributions of orbital elements we hope to learn something about the typical
dynamical configuration when stars form ({\it inverse dynamical population
synthesis}).

This paper is the first in a series of three papers which we refer to as K1
(this paper), K2 (Kroupa 1995a) and K3 (Kroupa 1995b). In K2 we discuss the
properties of the model Galactic field star population under the hypothesis
that the stars originate in the dominant mode cluster, and elaborate on the
concept of eigenevolution. In K3 we discuss the observational
consequences of the dynamical evolution of star-forming aggregates, making use
of the simulations performed in papers K1 and K2.

The observational constraints on orbital parameters are discussed in Section~2.
We introduce our models in Section~3. In Section~4 we
perform inverse dynamical population
synthesis to isolate a possible dominant mode
of star-formation that may lead to the population
of Galactic field stars. By correcting the main sequence distribution of
orbital elements for the dynamical evolution in the dominant mode cluster
we derive an initial period distribution in Section~5. In Section~6 we discuss
our findings and Section~7 presents our conclusions.

\bigskip
\bigbreak
\noindent{\bf 2 THE OBSERVATIONAL CONSTRAINTS ON ORBITAL PARAMETERS}
\nobreak\vskip 10pt\nobreak
\noindent
In Fig.~1 we reproduce the distribution of periods, $f_{\rm P}$, for G, K~and
M~dwarf
systems. It is evident that $f_{\rm P}$ does not significantly depend on
spectral type. The total
proportion of late-type binary systems in the Galactic disk amounts to $f_{\rm
tot}^{\rm obs}=0.47\pm0.05$, which is a weighted average of the proportion of
binaries among~G, K~and M~dwarfs, respectively: $f_{\rm G}=0.53\pm0.08$,
$f_{\rm K}=0.45\pm0.07$ and $f_{\rm M}=0.42\pm0.09$ (see Leinert et al. 1993
and Fischer \& Marcy 1992 for the compilation of these numbers -- note that a
weighted average is not a strictly correct estimate of $f_{\rm tot}$, but
suffices for our purpose; $f_{\rm tot}$ is defined by equation~2 below, see
also equation~7). In the top panel
of Fig.~2 we show the mass-ratio distribution for binary systems with a solar
mass primary (Duquennoy \& Mayor
1991). We correct this mass-ratio distribution for the bias that
hampers spectroscopic surveys in that for a given orbital mass function
not all inclinations are accessible to an observer (Mazeh
\& Goldberg 1992). To this end we add
the long-period (log$_{10}P>3.5$) mass ratio distribution of Duquennoy \& Mayor
(1991) and the corrected short period (log$_{10}P<3.5$) mass ratio distribution
of Mazeh et al. (1992, equations~1 and~2).
The corrected
mass ratio distribution is shown in the
top panel of Fig.~2 as well as the expected mass ratio
distribution
for binaries with a solar mass primary if the secondaries were to follow the
Galactic field star mass function. The G-dwarf binary star data show a
depletion of small mass ratio systems. This discrepancy has also been pointed
out by Kroupa \& Tout (1992). In the middle panel of Fig.~2 we contrast
the mass-ratio distribution for short-period systems (Mazeh et al. 1992)
with the distribution for long-period systems (Duquennoy \& Mayor 1991) after
normalising both to unit area. It appears that both are fundamentaly different.
It is noteworthy that Goldberg \& Mazeh (1994) find a mass ratio distribution
for short period Pleiades binaries rather similar to that shown here,
although the statistics is very poor. The CORAVEL study also shows that
the eccentricities for G-type main sequence binaries with periods less than
approximately 12~days are circularised. For $11.6<P<1000\,$days the
eccentricity
distribution is bell shaped with a mean eccentricity $\overline{e}=0.31\pm0.04$
whereas for $P>1000\,$days Duquennoy \& Mayor find the data are consistent with
the thermal distribution (Section~3.2.1). These eccentricity distributions are
shown in the bottom panel of Fig.~2. Both
the mass-ratio as well as the eccentricity distributions undergo a significant
change of character at log$_{10}P\approx3$.  This point will be further
discussed in K2.

The data that are currently available for pre-main sequence binaries are still
somewhat scarce (for a more in depth review of the observations see Mathieu
1994). In Fig.~1 we
compile the measured distributions in log$_{10}P$ obtained from the
separations of the components. The histograms for pre-main sequence binaries
shown in Fig.~1 are obtained from Simon (1992), Leinert et al. (1993) and
Richichi et al. (1994), and by rebinning the data
assuming a mean stellar mass of $1.4\,M_\odot$, $1\,M_\odot$ and $1.4\,M_\odot$
published by Reipurth \& Zinnecker (1993) and Ghez et al.
(1993), respectively, into
$\Delta {\rm log}_{10}P=1$ wide bins.  Mathieu (1992) finds that the proportion
of short period pre-main sequence binaries (log$_{10}P<2$) is virtually the
same as for main sequence binaries. Leinert et al. (1993) discuss the
distribution of brightness ratios for their sample of pre-main sequence
binaries. Although the mapping to a mass ratio distribution is very uncertain
they find that their data are consistent with random pairings from the
KTG(1.3) mass function, which increases as a power-law with
decreasing mass (equation~1 below). This mass function provides a better fit to
their data than a mass function which flattens or decreases below about
$0.3\,M_\odot$, and which fits the mass-ratio distribution of
solar-type binaries (Duquennoy \& Mayor 1991).

Comparison of the period distribution of pre-main sequence binaries
with the period distribution of
main sequence binaries with a G, K and M star primary suggests
evolution
of the distributions. Comparing the main sequence data with the pre-main
sequence data shown in Fig.~1 we must keep in mind that triple and
quadruple systems
are counted by Duquennoy \& Mayor (1991) as two and three orbits, respectively,
in the former and that the latter
data are derived from apparent separations on the sky. However, the discussion
in this paper is not affected by the different nature of the two
samples.

The pre-main sequence binary star data were obtained from observations of
distributed star forming regions (Taurus--Auriga and Ophiuchus--Scorpius).
Individual
star-forming regions might lead to distributions of orbital parameters that
differ from each other, as discussed by Reipurth \& Zinnecker (1993),
Leinert et al. (1993),  Ghez et al. (1993),
Mathieu (1992), Simon (1992) and Durisen \& Sterzik (1994). However, variation
of the binary star distribution at birth between different star-forming regions
is suggested but by no means established by the data at present. Furthermore,
we stress
that even if different binary proportions and period distributions are observed
in different star forming regions it must first be established whether the
differences are not due to different dynamical ages of the regions rather than
being the result of different star-formation conditions, as we discuss in
greater length in Section~6.3 and in K3.

\bigskip
\bigbreak
\noindent{\bf 3 MODELS}
\nobreak\vskip 10pt\nobreak
\noindent In this section a brief description of the numerical tool is given,
and the set of numerical experiments perfomed in Section~4 are
detailed.
\nobreak\vskip 10pt\nobreak
\noindent{\bf 3.1 The N-body program}
\nobreak\vskip 10pt\nobreak
\noindent
We need to integrate stellar orbits in a system with a few hundred binary
stars with the principal aim of studying the evolution of their orbital
elements. The processes responsible for the change of orbital parameters of a
binary
system operate on a timescale that is many orders of magnitudes less than the
timescale over which an aggregate of stars evolves. Both timescales are
important for our work.

Aarseth has been developing a N-body code for the direct integration
of the orbits of stars in a cluster (Aarseth
1994, see also Aarseth 1992). We make use of his {\it nbody5} code
in which the regularisation of the equations of motion of more than two
point
stars in close proximity (Mikkola \& Aarseth 1990, 1993) makes the
computational task we are faced with possible. We model a standard Galactic
tidal field, with Oort's constants $A=14.4$~km~sec$^{-1}$~kpc$^{-1}$, and
$B=-12$~km~sec$^{-1}$~kpc$^{-1}$, and a local mass density of
$0.11\,M_\odot$~pc$^{-3}$.
We retain all stars during the simulations
irrespective of whether stars are bound to the aggregate or not

To identify all bound binary systems we supplement {\it nbody5} by an
extensive data reduction program which searches all position and velocity
vectors (with respect to the local standard of rest) of the stars output by
{\it nbody5} at some prechosen time intervals to find the binaries.
Each star
$i$ is paired with every other star $j$ to find that pair for star $i$ with the
smallest separation of components. All pairs $i$ thus found are compared to
exclude repetition of components. The binding energy for each surviving pair is
computed, and if negative, this pair is classified as a bound binary system,
for which orbital elements are computed.

The orbital
elements thus found need to be checked for stability against numerically
induced evolution. This is achieved by comparing the eccentricity--period data
at different time intervals for an aggregate with half mass radius of
$31\,$pc
consisting of two hundred binaries. Such a large half mass radius ensures that
the inter-binary distance is large so that stimulated evolution of the
orbital parameters is negligible. Binary systems with periods smaller than
$10^8$days show no changes in their orbital parameters over a time span of
$1.3\,$Gyr. Systems with periods larger than this are subject to perturbation
by the Galactic tidal field and suffer changes in eccentricity that increase
with increasing period. Systems that initially had a period larger than
approximately $10^{10.3}$days (corresponding to a semi-major axis
of approximately $0.6\,$pc for a system mass of $0.64\,M_\odot$) are
disrupted. The widest binary systems known have a separation of components of
$0.1\,$pc (Close, Richer \& Crabtree 1990) and we are thus satisfied that the
N-body
program correctly integrates the motion of stars in isolated binaries.

\nobreak\vskip 10pt\nobreak
\noindent{\bf 3.2 Initial parameters}
\vskip 10pt\nobreak
\noindent{\bf 3.2.1 Binary stars}
\vskip 10pt\nobreak
\noindent
We concentrate on low-mass stars with a mass between
0.1 and $1.1\,M_\odot$ because these make up the bulk of the stellar
population in the Galactic disk. It is for these that we have the
best measurements of binary star distribution functions and we do not have to
take into account post-main sequence evolution, when mass loss through stellar
winds and supernova explosions complicate the dynamics of a stellar aggregate.
Also, the recent studies on
dublicity in star-forming regions concentrate on T-Tauri type stars. The
masses are chosen from the following mass-generating function (Kroupa et al.
1993):

$$m(X) = 0.08 + {\gamma_1\,X^{\gamma_2} + \gamma_3\,X^{\gamma_4}
\over (1-X)^{0.58}}, \eqno (1a)$$

\noindent where $X$ is distributed uniformly in the range 0~to~1, $m$ is the
stellar mass in solar units and
$\gamma_1=0.19$, $\gamma_2=1.55$,
$\gamma_3=0.050$ and $\gamma_4=0.6$ if $\alpha_1=1.3$ in equation~1b.
Equation~1 leads to a stellar mass spectrum in the local Galactic disk, derived
from both nearby and deep star count data. After correcting for post-main
sequence stellar evolution it can be approximated conveniently by a three-part
power law initial mass function:


$$\xi (m) = 0.087\cases{{\displaystyle 0.5^{\alpha_1}\,m^{-\alpha_1}},
&if $0.08 \le m < 0.5$; \cr
{\displaystyle 0.5^{2.2}\,m^{-2.2}},
&if $0.5 \le m < 1.0$, \cr
{\displaystyle 0.5^{2.2}\,m^{-2.7}},
&if $1.0 \le m < \infty$, \cr}\eqno (1b)$$

\noindent where $\xi(m)\,dm$ is the number of stars per pc$^3$ in the mass
range $m$ to $m+dm$. We refer to this spectrum of stellar masses as the
KTG($\alpha_1$) mass function and adopt $\alpha_1=1.3$.
The mean stellar mass of our population is $0.32\,M_\odot$. Stars more massive
than $1.1\,M_\odot$ contribute 8~per cent by number and~35 (39)~per cent by
mass to the stellar population if the most massive star born has a mass of~10
(50)~$M_\odot$.

A critical analysis of the methods used to derive mass-ratio distributions by
Tout (1991) suggests that the simplest of selection effects can account for
most structure claimed in the observed mass-ratio distribution.
The lack of evidence for significant correlation is further
discussed by Kroupa et al. (1993). As discussed in the
introduction, Leinert et al. (1993) find evidence that the component masses
of pre-main sequence stars are chosen at random from equation~1.
Consequently we pair the stellar
masses generated from equation~1a at random to form a binary star population
that has an uncorrelated mass ratio distribution $f_{\rm q}(q)$ at birth
(shown in fig.~12 in K2), where $f_{\rm q}(q)\,dq$ is the proportion
of systems with a mass ratio in the range $q$ to $q+dq$. Here $q=m_2/m_1\le1$,
where $m_1$ and $m_2$ are the mass of the primary and secondary component,
respectively.

The initial orbital parameters of a freshly hatched binary star population are
completely unknown.
Because the binary proportion is high on the main sequence and even higher
in star-forming regions we assume that all stars form as binaries, {\it i.e.}
we set $f_{\rm tot}=1$ initially. We define the overall fraction
of binaries at time $t$ as

$$f_{\rm tot}(t) = {N_{\rm bin}(t) \over{N_{\rm sing}(t) + N_{\rm bin}(t)}},
\eqno (2)$$

\noindent where $N_{\rm bin}(t)$ and $N_{\rm sing}(t)$ are the number of
multiple
systems and single stars, respectively, and the denominator is the number of
all systems
({\it c.f.} Reipurth \& Zinnecker 1993).

The distribution function of semi-major axes, $a$, is approximated rather well
by a Gaussian in log$_{10}a$ for binaries on the main sequence (Fig.~1). Our
assumption here is that the initial distribution is flat, straddling the peak
at $a\approx30$~AU
in the main-sequence distribution of semi-major axes.
 The generating function is


$$ a(X) = {a_{\rm min}\, 10^{X\,{\rm log}_{10}\left({a_{\rm max}\over
a_{\rm min}}\right)}}, \eqno (3a)$$

\noindent where $a_{\max}$ and $a_{\rm min}$ are the maximum and minimum
semi-major axes we wish to include and $X\epsilon$[0,1] is a uniform random
variate. We choose $a_{\rm max}=1690\,$AU and $a_{\rm min}=1.69\,$AU.
The distribution function is

$$ f_{\rm a}({\rm log}_{10}a) = \left[{\rm log}_{10}(a_{\rm max}) - {\rm
log}_{10}(a_{\rm min})\right]^{-1}, \eqno (3b)$$

\noindent where $dn({\rm log}_{10}a) = N_{\rm bin}\,f_{\rm a}({\rm
log}_{10}a)\,d{\rm
log}_{10}a$ is the number of orbits with semi-major axis in the range
log$_{10}a$ to log$_{10}a+d{\rm log}_{10}a$.
This distribution is
equivalent to a flat log$_{10}P$
distribution with $10^3\le {\rm log}_{10}P \le 10^{7.5}$ (for a mean system
mass of $0.64\,M_\odot$) which is consistent with the pre-main sequence data
(Fig.~1).

We assume the initial eccentricity distribution function represents a
population in statistical equilibrium.
The generating function is

$$ e(X) = \sqrt{X}, \eqno (4a)$$

\noindent which has the thermal distribution function

$$ f_{\rm e}(e) = 2\,e. \eqno (4b)$$

\noindent The
number of systems with eccentricity
in the range $e$ to $e+de$ is $dn(e)=N_{\rm bin}\,f_{\rm e}(e)\,de$ (see for
example Heggie 1975). This dynamically relaxed distribution has the advantage
that most
binary systems have large eccentricity in agreement with recent studies of star
formation (Boss 1988, Pringle 1989, Bonnell \& Bastien 1992, Elmegreen 1993,
Burkert \& Bodenheimer 1993), although theory cannot predict orbital elements
because computations have to be stopped before the stars have accreted most of
the natal material.  The results of this paper are not sensitive to the form of
$f_{\rm e}(e)$ (see K2).

In summary, we assume the initial distribution function can be separated:

$$ D(P,e,q) = f_{\rm P}({\rm log}_{10}P)\times f_{\rm q}(q)\times f_{\rm e}(e),
\eqno (5)$$

\noindent where $f_{\rm P}$ is the distribution of log$_{10}P$ and is obtained
from equation~(3b) using Kepler's equation.

\vskip 10pt\nobreak
\noindent{\bf 3.2.2 The stellar aggregates}
\vskip 10pt\nobreak
\noindent
Stars in the Galaxy are formed in groups in molecular clouds but it is not
clear which parameters of these groups are typical for field stars. In
Section~4 we model the evolution of a number of initial aggregates of stars to
later
identify those that lead most closely to the distribution of binary properties
observed on the main sequence.

To cover a range of
possibilities we adopt Plummer models with various radii as given in Table~1.
Generating functions for
Plummer spheres can be found in Aarseth, Henon \& Wielen (1974).
We assume the aggregates are in
virial equilibrium, i.e. we choose $Q=-K/W=0.5$, where $W$ is the total binding
energy of the cluster and $K$ is the sum of the kinetic energies of all centre
of masses in the cluster. In our simulations stars are treated as point
particles.

We emphasise that we choose Plummer models merely for our convenience and that
we do not insist that stars form in such systems. Our choice of aggregate
can only be a first try. We must keep in mind that the early internal
dynamics of a cluster of proto stars is dominated by a rapidly changing
potential because after the stars begin to form a large proportion of the mass,
which remains as gas and dust, of most star-forming systems is removed
within of the order of $10^7$~years (Mathieu 1986, Battinelli \&
Capuzzo-Dolcetta 1991).
Most groups of young stars may become gravitationally unbound on
this timescale (we return to this point in Section~6.4). Also, we expect that
about 8~per~cent of all stars have masses
larger than $1.1\,M_\odot$. By excluding these and the changing background
potential we introduce a stellar mass dependent bias because mass
segregation leads to G~dwarfs (which are the most massive stars in our
simulations) spending, on average, more time near the cluster
centre and are thus subject to stimulated evolution over a longer time scale
than the less massive stars. The dynamics in a real cluster of young stars
is affected by mass loss from evolving stars and supernovae. Finally, the
finite size of stars leads to (rare) stellar collisions and mergers, as well as
removal of kinetic energy through tidal dissipation in close fly-by events.
These
complications do not, however, affect the results presented in
this paper.

In Table~1, Column~1 lists the paper which discusses the simulations (K1$=$this
paper, K2$=$Kroupa 1995a, K3$=$Kroupa 1995b).
Column~2 lists the initial half mass radius, $R_{\rm
0.5}$. We choose four initial values of $R_{0.5}$ which span the range 0.077~pc
to 2.53~pc. The lower limit corresponds to tightly clustered star formation
(e.g. Trapezium Cluster), and the upper limit approximates isolated star
formation (e.g. Taurus--Auriga). Columns~3 and~4 contain the initial number of
binary
systems and the number of systems composed of single stars, respectively.
The aggregate mass is in all cases $128\,M_\odot$. We do not expect that each
star-forming event produces exactly 400~stars, but we consider this assumption
to be an adequate first guess. In this context it is of interest to note that
Battinelli, Brandimarti \& Capuzzo-Dolcetta (1994) find that the mass function
of open clusters peaks at a mass of $126\,M_\odot$, which is consistent with
the IRAS selected sample of embedded clusters obtained by Carpenter et al.
(1993).
Columns~5, 6~and~7 contain, respectively, the initial and final proportion of
binaries and the initial number density $n$
of stars within a sphere with a radius of $2\,$pc centred on the number
density maximum, $n_{\rm c}$, the initial value of which is listed in Column~8.
Our maximum central number density is comparable to that observed in the
Trapezium Cluster ($\approx4.7\times10^4$ stars~pc$^{-3}$, McCaughrean \&
Stauffer 1994). Our smallest central number density is an adequate
approximation of isolated star formation (e.g. Taurus--Auriga).
Column~9 gives the initial average
velocity dispersion for each system and Columns~10 and~11 list the initial
crossing time, $T_{\rm cr}=2\,R_{0.5}/\sigma$,
and median relaxation time,
$T_{\rm relax}\propto R_{0.5}^{3\over2}N_{\rm bin}^{1\over2}/{\rm
log}_{10}(0.8\,N_{\rm bin})$ (Spitzer \& Hart
1971), respectively. The number of simulations per cluster
($\ge3$ to increase statistical significance) with
different initial random number seeds
are given in Column~12. In all subsequent discussions the mean values of the
$N_{\rm run}$ runs are used for the relevant parameters, and their quoted
uncertainties are standard
deviations of the mean. The mean lifetime, $\tau_{0.1}$, of each cluster
is given in Column~13 and is defined in~K3. It is the time required for
the central number density to drop below 0.1~stars~pc$^{-3}$, and signifies
complete disintegration. We shall refer to `final' distributions as
distributions evaluated after aggregate disintegration at $t=1\,$Gyr. Finally,
in Column~14 we list the
ratio of the initial cluster binding energy, $W$, and of the binding energy of
a binary, $E_{\rm b}$, with $1\,M_\odot$
companions and shortest period occurring in the birth period distributions
used (i.e. in K2 prior to eigenevolution).
The single-star runs ($N_{\rm bin}=0$), which are of academic interest only,
are used for comparison with the evolution of the realistic aggregates
consisting initially only of binaries.

\bigbreak
\vskip 3mm
\bigbreak

\hang{ {\bf Table 1.} Initial stellar aggregates studied in papers p.K1, K2
and K3

\nobreak
\vskip 1mm
\nobreak
{\hsize 17 cm \settabs 21 \columns

\+p.K &$R_{\rm 0.5}$ &$N_{\rm bin}$ &$N_{\rm sing}$ &~~$f_{\rm tot}$
&~~~~~~$f_{\rm tot}$
&&&$n$~~~log$_{10}(n_{\rm c})$& &&$\sigma$
&$T_{\rm cr}$ &&$T_{\rm relax}$ &$N_{\rm run}$ &~~~$\tau_{0.1}$ &&&$E_{\rm
b}/W$
\cr
\+ &pc & & &~$t=0$ &~~~$t=1$~Gyr
&&&stars &stars  &&km
&Myrs &&Myrs & &~~~Myrs \cr
\+&&&&& &&&pc$^{-3}$ &pc$^{-3}$&&sec$^{-1}$ \cr

\+1,3 &0.08 &200 &~~0 &~~~1&$0.269\pm0.015$ &&&12 &5.6 &&1.7 &0.094 &&0.30
&5
&$655\pm137$
&&&0.7\cr
\+1,3 &0.25 &200 &~~0 &~~~1&$0.400\pm0.020$ &&&12 &4.1 &&0.9 &0.54  &&1.8
&5
&$696\pm135$
&&&2\cr
\+1,3 &0.77 &200 &~~0 &~~~1&$0.610\pm0.037$ &&&11 &2.7 &&0.5 &3.0   &&9.5
&5
&$748\pm114$
&&&7\cr
\+1,3 &2.53 &200 &~~0 &~~~1&$0.823\pm0.018$ &&&5  &1.1 &&0.3 &17    &&56
&5
&$693\pm127$
&&&20\cr
\+1,3 &0.08 &0 &~~400 &~~~0&$0.021\pm0.006$ &&&12 &5.6 &&1.7 &0.094 &&0.30
&3
&$736\pm178$
&&&--\cr
\+1,3 &0.25 &0 &~~400 &~~~0&$0.020\pm0.006$ &&&12 &4.1 &&0.9 &0.54  &&1.8
&3
&$732\pm138$
&&&--\cr
\+1 &0.77 &200 &~~0 &~~~1&$0.553\pm0.048$ &&&11 &2.7 &&0.5 &2.9   &&9.5
&5
&$705\pm247$
&&&490\cr
\+1 &0.77 &200 &~~0 &~~~1&$0.416\pm0.042$ &&&11 &2.7 &&0.5 &2.9   &&9.5
&5
&$686\pm85$
&&&490\cr

\+2,3   &0.85 &200 &~~0 &~~~1&$0.480\pm0.032$ &&&10 &2.5 &&0.5 &3.5   &&11
&20
&$740\pm150$
&&&120\cr
}
\bigbreak\vskip 3mm

\bigskip
\bigbreak
\noindent{\bf 4 INVERSE DYNAMICAL POPULATION SYNTHESIS}
\nobreak\vskip 10pt\nobreak
\noindent In this section we show that the dynamical properties of stellar
systems in the Galactic field can be derived if the majority of stars form in a
characteristic aggregate.

\nobreak\vskip 10pt\nobreak
\noindent{\bf 4.1 The dynamical properties of stellar systems}
\nobreak\vskip 10pt\nobreak
\noindent Assume we know that star formation produces stellar systems with a
distribution of birth {\it dynamical properties},
${\cal D}(\xi(m), D(P,e,q), N,
R_{0.5})$, which is the distribution of initial mass functions, of period,
eccentricity and mass-ratio distributions and of the number of stars forming in
a volume characterised by some half-mass radius. The Galactic field
population is given by the time evolved integral of ${\cal D}$ over all star
forming
events ever to have occurred. Thus, if we know ${\cal D}$, we can compute the
dynamical properties (stellar masses, orbital parameters,
multiplicities)
of the stellar population of the Galactic field under some assumed star
formation history.

However, ${\cal D}$ is completely unknown. We invert the problem here by
investigating if a representative set $\xi(m), D(P,e,q), N$ and $R_{0.5}$ which
represents the most common mode, or average of ${\cal D}$, can be deduced
from the currently observed stellar dynamical properties
in the Galactic disc. Kroupa et al. (1993)
have determined $\xi(m)$ (equation~1) and in Section~3.2.1 we adopt
a $D(P,e,q)$ which is consistent with all available pre-main sequence data. In
Section~3.2.2 we argue that $N_{\rm bin}=200$ is a reasonable first
guess, and we are left to estimate a characteristic, or representative,
$R_{0.5}$.

\nobreak\vskip 10pt\nobreak
\noindent{\bf 4.2 Stimulated evolution of orbital paramters of binary
stars}
\nobreak\vskip 10pt\nobreak
\noindent
As mentioned in the introduction, past work has etablished that binary systems
are ionised at a rate which is a function of the ratio of the internal binding
energy of the binary and the kinetic energy, or temperature, of the
surrounding field population. In fig.~5 of K3 we discuss the initial and
final distributions of binary star binding energies and centre of mass kinetic
energies for each aggregate. Preferentially those binary systems are ionised
that are relatively least bound.

The overall evolution of the binary star population is best summarised by
studying the evolution of the proportion of binaries, $f_{\rm tot}(t)$, given
by equation~2. In Fig.~3 we plot $f_{\rm tot}$ as a function of time, $t$,
for the first four aggregates listed
in Table~1. We also show $f_{\rm tot}$ for the additional simulations with no
primordial
binaries ($N_{\rm bin}=0$). From Fig.~3 we deduce that the Galactic disk main
sequence stellar population,
for which $f_{\rm tot}^{\rm obs}=0.47\pm0.05$, is best reproduced if
$0.25\,{\rm pc}<R_{0.5}<0.8\,$pc. Three further points are
displayed by Fig.~3: (i) The rate of evolution of $f_{\rm tot}$, and for that
matter of
all binary star orbital parameters, is largest while the cluster is dynamically
young, but proceeds much slower in terms of absolute time in clusters with
larger initial size. Thus, for example, we expect a very young (one Myr old)
embedded cluster with high central number density (about
$10^5$~stars~pc$^{-3}$) to have a significantly depleted binary star
population.  An example of such a cluster is the Trapezium Cluster which we
will discuss in greater detail in K3.
(ii) The final proportion of binaries is smaller when the
initial number density is larger. The reduction of the proportion of multiple
systems with increasing number of systems per cloud suggested by fig.4 of
Reipurth \& Zinnecker (1993) is interesting in this context.
(iii) The proportion of binaries formed by capture
($f_{\rm tot}\approx0.02$) is insignificant compared to the observed proportion
($f_{\rm tot}^{\rm obs}=0.47\pm0.05$) which is merely confirmation of the
long established
understanding that dissipationless capture is not a viable formation mechanism
for binary systems (see Bodenheimer et al. 1993 and references therein), even
if initial subclustering, which enhances the proportion of binaries formed by
capture (Aarseth \& Hills 1972), is taken into account.

The binding energy of a binary system can be written

$$-E_{\rm bin} = -G\,{m_1^2 \over 2\,a}\,q.  \eqno (6)$$

\noindent We thus have $E_{\rm bin}\propto q$ and we
immediately see that in a cluster the mass ratio distribution must be depleted
such that systems with small $q$ are ionised at a larger rate than those with
$q\approx1$. In Fig.~4 we show the mass function of secondaries in systems
with a primary mass
$0.85\,M_\odot < m_1 < 1.1\,M_\odot$ for each of the first four aggregates
listed in Table~1 initially
and after aggregate dissolution. The increasing bias towards $q=1$
with decreasing initial aggregate size is apparent, and comparison with the
G-dwarf main sequence binary star distribution suggests that the observed
distribution is
best reproduced if $R_{0.5}\approx0.8\,$pc.

The distribution of periods is expected to be more depleted at long
periods than at short periods. In Fig.~5 we show the initial and
final distributions of periods of our binary star population.
Normalisation of the histograms is obtained from

$$ f_{\rm P,i}(t) = {N_{\rm bin,i}(t) \over N_{\rm bin}(t) + N_{\rm sing}(t)}
\eqno (7a)$$

\noindent with

$$\sum_{\rm all~bins~i} f_{\rm P,i}(t) = f_{\rm tot}(t),  \eqno (7b)$$

\noindent where $N_{\rm bin,i}$ is the number of orbits in the ith
log$_{10}P$ bin,
and the denominator is as in equation~2. On comparison with the observed period
distribution (Fig.~5) we deduce that the main
sequence period distribution is best reproduced if $R_{0.5}\approx0.8\,$pc.
{}From Fig.~5 we also see that the number of orbits with log$_{10}P<3$ and
log$_{10}P>7.5$ does not increase significantly in any of the aggregates. This
lack of hardening and softening of binaries comes about because the lifetimes
of the aggregates studied here are too short ({\it total} disintegration after
about 700~Myrs, see K3).

The eccentricity distribution remains thermal (equation~4) at all times and in
all clusters.

\nobreak\vskip 10pt\nobreak
\noindent{\bf 4.3 Dynamically equivalent aggregates}
\nobreak\vskip 10pt\nobreak
\noindent
Given our assumptions which are consistent with all available
observational constraints (Section~3.2.1) we have shown that the dynamical
properties
of Galactic field systems can be derived if most stars form in an aggregate
with $(N_{\rm bin},R_{0.5})\approx(200,0.8\,{\rm pc})$.

We postulate that most
stars may have formed in aggregates which are dynamically equivalent to
this {\it dominant mode cluster}.

We define a stellar aggregate or cluster to be {\it dynamically equivalent} to
the dominant mode cluster if stimulated evolution evolves the initial
population of stellar systems (with initial dynamical properties as defined in
Section~3.2.1 and equation~11 below) to a population which has final stellar
dynamical properties that are consistent with the observed properties of
Galactic field stellar systems (see also section~4.2 in K3).

\bigskip
\bigbreak
\noindent{\bf 5 THE INITIAL PERIOD DISTRIBUTION}
\nobreak\vskip 10pt\nobreak
\noindent
Star formation theory is not currently in a state that allows prediction of
either the proportion of multiple systems formed, or the distribution of
periods of the primordial binary star population (see e.g. Bonnell 1994).
Any clues pertaining to the primordial orbital parameters of a
binary star population are thus very valuable. In Section~4.2 we have
seen that even very simple models of the initial period, eccentricity and
mass-ratio distribution (which are consistent with the available
observational data) lead to encouraging
results given they are subject to the internal dynamics of a stellar aggregate.

Investigating Fig.~5 we conclude that it must be possible,
by correcting the observed main sequence period distribution for
stimulated evolution, to identify an
initial period distribution that evolves to the observed main sequence
distribution. We proceed as follows: The simulations discussed in Section~4.2
are our zeroth order iteration. In Sections~5.1 and~5.2 we go to the first and
second order iteration, respectively which are the
likely range of the initial period distribution.

\nobreak\vskip 10pt\nobreak
\noindent{\bf 5.1 The first iteration}
\nobreak\vskip 10pt\nobreak
\noindent
We calculate the orbit depletion function

$$ c_{\rm P,i}(t) = {f_{\rm P,i}(t) \over f_{\rm P,i}(0)}, \eqno (8)$$

\noindent where $f_{\rm P,i}(t)$ is the proportion of orbits in the i$^{\rm
th}$ log$_{10}P$ bin at time $t$ (equation~7). The ratio $c_{\rm P,i}(t)$
measures the depletion of orbits in the i$^{\rm th}$ period bin. The final
distribution is
plotted in Fig.~6. Hardened binaries now appear at log$_{10}P<3$ ($c_{\rm
P,i}>1$). If we denote the observed main sequence period distribution (Fig.~1)
by $f_{\rm P,ms,i}$ then the initial period distribution is given by

$$ f_{\rm P,in,i} = f_{\rm P,ms,i} \,c_{\rm P,i}(1\,{\rm Gyr})^{-1} \eqno
(9a)$$

\noindent with the constraint

$$ \int_{{\rm log}_{10}P_{\rm min}}
^{{\rm log}_{10}P_{\rm max}} f_{\rm P,in}\,d{\rm log}_{10}P = 1
\eqno (9b)$$

\noindent which becomes a sum for the discrete case (equation~9a).
In Fig.~7 $f_{\rm P,in,i}$ is plotted for the first four aggregates listed
in Table~1. From Fig.~7 we deduce that $f_{\rm P,in,i}\propto
{\rm
log}_{10}P$ approximately. We adopt as our first try the distribution function

$$ f_{\rm P,in}^{(1)} = \beta\,\left[{\rm log}_{10}(P)-{\rm
log}_{10}(P_{\rm min})\right] \eqno (10a )$$

\noindent with the period generating function

$${\rm log}_{10}P(X) = ({2\over\beta}X)^{1\over2} + {\rm log}_{10}P_{\rm min}
\eqno (10b)$$

\noindent where $X\epsilon[0,1]$ is a uniform random variate.
Fig.~7 suggests log$_{10}P_{\rm min}\approx 2$. However,
this choice would imply a deficit in main sequence orbits for
log$_{10}P<3$. This deficit of orbits would not be made up by stimulated
evolution as hardening of binary orbits is not efficient enough
(Section~4 and K2).
We address the question as to whether there is a $P_{\rm
min}>1\,$day in~K2 and for the present adopt log$_{10}P_{\rm
min}=0$.

The slope $\beta$ and maximum period $P_{\rm max} (X=1)$ are
tabulated in Table~2. With our
above choice for log$_{10}P_{\rm min}$ the linear approximation given by
equation~(10a) only allows a rough measurment of $\beta$ for the aggegates with
$R_{0.5}=0.08, 0.25, 2.53\,$pc (see Fig.~7).

\bigbreak
\vskip 3mm
\bigbreak

\hang{ {\bf Table 2.} Preliminary initial period distribution (equation~10)

\nobreak
\vskip 1mm
\nobreak
{\hsize 15 cm \settabs 5 \columns

\+& $R_{\rm 0.5}$ &$\beta$ &log$_{10}P_{\rm max}$
\cr
\+&(pc)  & &(days) \cr
\+\cr
\+&0.08 &$\approx0.058$ &5.9 \cr
\+&0.25 &$\approx0.042$ &6.9 \cr
\+&0.77 &~$0.034$      &7.67 \cr
\+&2.53 &$\approx0.020$ &10 \cr

}
\bigbreak\vskip 3mm

\noindent From Table~2 we deduce that aggregates approximately with
$R_{0.5}\ge0.77\,$pc
require $\beta$ that imply $P_{\rm max}>10^{7.5}$~days which is a {\it
necessary} condition for
agreement with the observational data. Also, from our discussion of the mass
ratio distribution in Section~4.2 we require $R_{0.5}<2.5\,$pc and we now adopt
$R_{0.5}=0.77\,$pc, $\beta=0.034$ and $P_{\rm max}=10^{7.67}\,$days as
best approximating all observational constraints.

Thus we find that in order to be able to model the full range of orbital
periods observed and the observed mass ratio distribution of long-period
G~dwarf binaries we require $R_{0.5}\approx0.8\,$pc.

We perform $N_{\rm run}=5$ runs of an aggregate with $R_{0.5}=0.77\,$pc (the
first cluster listed under p.K1 in Table~1).
The final period distribution is shown in Fig.~8a.
Comparing with the main sequence data we find that at
it is somewhat too large for log$_{10}P>5$
and falls too steeply for log$_{10}P>7$. The final proportion of binary systems
is $f_{\rm tot}^{(1)} = 0.55\pm0.05$ (cf. $f_{\rm tot}^{\rm obs}=0.47\pm0.05$).

It is highly improbable that the initial period distribution does in fact have
the form of equation~10 although we cannot reject the resulting model main
sequence
distribution with high confidence. In particular, it is worthwhile to point
out that Close et al. (1990) compiled a complete sample of wide
binaries
in the solar neighbourhood and found systems with separation of
components as large as 0.1~pc. This is equivalent to log$_{10}P=7.5$ if the
system mass is $1.4\,M_\odot$. A power law decay of $f_{\rm P,ms}$ with
increasing period cannot be rejected though, and Duquennoy \& Mayor (1991)
find evidence that
systems with periods as large as $10^9\,$days exist. As equation~10 implies a
sharp fall-off of the simulated $f_{\rm P,ms}$ for log$_{10}P>7$ we expect the
initial period distribution, $f_{\rm P,in}$,
must have log$_{10}P_{\rm max}>7.67$. This requires a flattening at large
periods given constraint~(9b).

\nobreak\vskip 10pt\nobreak
\noindent{\bf 5.2 The second iteration}
\nobreak\vskip 10pt\nobreak
\noindent A distribution function which fulfills the requirement raised at the
end of Section~5.1 and which is also easily integrable is

$$ f_{\rm P,in}^{(2)} = \eta \, { \left({\rm log}_{10}P -
{\rm log}_{10}P_{\rm min}\right)  \over \delta + \left({\rm log}_{10}P -
{\rm log}_{10}P_{\rm min}\right)^2}  \eqno (11a)$$

\noindent with the period generating function

$$ {\rm log}_{10}P(X) = {\rm log}_{10}P_{\rm min} +
\left[\delta\left(e^{2\,X\over\eta}-1\right)\right]^{1\over2}  \eqno(11b)$$

\noindent where $X\epsilon[0,1]$ is a uniform random variate. For our
second iteration we also assume log$_{10}P_{\rm min}=0$.
Adjustment of the parameters such that the slope of the distribution at small
periods is approximately the same as for $f_{\rm P,in}^{(1)}$ and
log$_{10}P_{\rm max}>7.5$ leads us to adopt $\eta=3.50$ and $\delta=100$ with
log$_{10}P_{\rm max}=8.78$ as our second iteration.

The results of five simulations of a cluster with $R_{0.5}=0.77\,$pc
(second cluster listed in Table~1 under p.K1) are shown in Fig.~8b. The final
proportion of binary systems is $f_{\rm tot}^{(2)}=0.42\pm0.04$ (cf. $f_{\rm
tot}^{\rm obs}=0.47\pm0.05$).

\nobreak\vskip 10pt\nobreak
\noindent{\bf 5.3 Result}
\nobreak\vskip 10pt\nobreak
\noindent
The comparison of Figs.~8a and~8b, and $f_{\rm tot}^{(1)}$ and $f_{\rm
tot}^{(2)}$ with $f_{\rm tot}^{\rm obs}$, shows that
$f_{\rm P,in}^{(1)}$ and $f_{\rm P,in}^{(2)}$ are probably close to being upper
and lower bounds of the initial period distribution.

We have found that
in order to be able to model the full range of orbital
periods observed
we require $R_{0.5}\approx0.8\,$pc, in agreement with our
conclusion in Section~4.

\nobreak\vskip 10pt\nobreak
\noindent{\bf 6 DISCUSSION}
\nobreak\vskip 10pt\nobreak
\noindent{\bf 6.1 Clustered star formation as the dominant mode of star
formation?}
\nobreak\vskip 10pt\nobreak
\noindent
Perhaps the most interesting insight we have gained here is that our inverse
dynamical population synthesis suggests that clustered star formation may be
the dominant mode rather than isolated or distributed star formation. We have
been able to obtain this insight by using the dynamical properties of observed
stellar systems in the Galactic disk as tracers of the evolutionary histories
of the systems, together with simple assumptions about the initial
dynamical properties that are consistent with all observational constraints.
Our finding is thus independent from and much more general than imaging surveys
of
star forming regions which are restricted to individual molecular clouds and
cannot at present detect pre-main sequence stars with masses lower than about
$0.5\,M_\odot$.

However, it is reassuring that surveys of various star forming regions find a
similar
result. Lada \& Lada (1991) list in their table~1 the physical characteristics
of the embedded clusters in the L1630 molecular cloud in the Orion complex. An
imaging survey at near-infrared wavelengths which can detect sources of a few
tenths solar mass found 627 sources. Of these 80~per cent are located in two
embedded clusters with radii of 0.86 and 0.88~pc, and the rest are located in
two embedded clusters with radii of 0.59 and 0.30~pc. Carpenter et al. (1993)
image 20 bright IRAS sources at near-infrared wavebands and find in all cases
that embedded clusters of low mass stars are associated with the IRAS sources
(usually one or two OB stars). Their radii are typically roughly 0.5~pc with
a typical stellar number density of 70~stars~pc$^{-3}$. These values are lower
limits because the faint stars are not detected, but demonstrate that these
specially selected systems are similar to our dominant mode cluster.
On the other hand, the near-infrared survey of the L1641 molecular cloud
(Strom, Strom \& Merrill 1993) finds a significant distributed population of
about 1500 about $5-7$~Myrs old stars, seven aggregates consisting of 10--50
stars
with volume densities of roughly $10^2$~stars~pc$^{-3}$ and ages less than
1~Myr, and one new embedded
cluster with about 150~stars and a volume density of $10^2$~stars~pc$^{-3}$.
Thus, direct imaging surveys cannot at present verify whether a single mode
(clustered or distributed) of star formation predominates. In addition, we
point out that the characteristics of the binary star population are unknown
for any of the above surveyed star-fomring systems. If our conclusions are
correct then the binary proportion must be very high in embedded clusters, and
the mass-ratio distribution of the young binaries must be approximately
uncorrelated. The details will be a function of the dynamical age of each
individual stellar aggregate.

Radial velocity and proper motion measurements of the distributed population
in L1641 may shed light on
the possibility that some or most of the distributed stars have migrated from
aggregates that are now dissolved. However, the spatial motion of a young star
after leaving the aggregate will probably change significantly because the
gravitational potential of the molecular cloud is not likely to be smooth. In
K3 we argue that a distributed population of young stellar objects is obtained
after dissolution of birth aggregates because most systems will have velocities
that are much smaller than the escape velocity from the molecular cloud. The
proportion of binaries and their orbital parameters is a potentially powerfull
discriminant for the origin of a distributed population (K3).

\nobreak\vskip 10pt\nobreak
\noindent{\bf 6.2 Concerning the stellar mass function...}
\nobreak\vskip 10pt\nobreak
\noindent
Our finding that most stars may have originated from aggregates that are
dynamically equivalent to our dominant mode cluster rests in part on our
assumed stellar mass function and the discrepancy with the mass-ratio
distribution of G~dwarf main sequence binary systems (Kroupa \& Tout 1992).
Although we are confident that
the KTG(1.3) mass function (equation~1) is a good approximation to the true
initial stellar mass function of Galactic field stars (Kroupa 1995c)
we now contemplate a different form to test the robusteness of our
conclusions.

For example the mass function could flatten below about
$0.5\,M_\odot$ or decrease below about $0.2\,M_\odot$, i.e. it could
have the
same form as the mass function of secondaries in G~dwarf binaries with
log$_{10}P>3$.
In this case we would have to argue that the stellar luminosity function for
nearby stars
does not reflect the true Galactic field luminosity function for single stars
(see Kroupa et al. 1993, Kroupa 1995c), and that the photometric luminosity
functions are the correct single star
luminosity functions, as was assumed for example by Kroupa, Tout \& Gilmore
(1990) in their initial analysis of structure in the mass--luminosity relation.

Ignoring this, and the strong exclusion of such a mass function by Kroupa et
al. (1993, their section~9), we would find that
our inverse dynamical population synthesis would identify as the most common
mode
of star formation an aggregate similar to the $R_{0.5}=2.53$~pc model listed in
Table~1 because in this case stimulated evolution must be kept to a minimum in
order not to
deplete the secondary mass function in G~dwarf binaries significantly (Fig.~4).
This mode corresponds roughly to the isolated or distributed mode in
star
formation with a central (i.e. maximum) initial density of approximately 13
stars~pc$^{-3}$.
However, we would then find a discrepancy with the observed proportion of main
sequence binaries ($f_{\rm tot}^{\rm obs}=0.47\pm0.05$, Fig.~3) unless only
about
50~per~cent of all systems are binaries at birth. This is in contradiction
with the observational evidence that the initial proportion of binaries is
close to unity, whereby we stress that this observational evidence is strictly
valid only for the distributed mode of star formation.
Thus, following this line of argument we arrive at a contradiction (see also
Section~6.5).

The KTG($\alpha_1$) mass function may have a different $\alpha_1$.
Star count data
constrain $\alpha_1$ to lie in the 95~per cent confidence range 0.70 -- 1.85
(Kroupa et al. 1993). Disregarding this finding for illustrative
purposes only and allowing
$\alpha_1>1.3$ our simulations would show that we need $R_{0.5}<0.8$~pc to
force the final mass-ratio distribution of solar-type binaries to have the same
shape as the observed distribution, as can be inferred by consulting Fig.~4. In
this case, however, we would ionise too many binary systems  implying $f_{\rm
tot}<f_{\rm tot}^{\rm obs}$. Alternatively, if $\alpha_1<1.3$ ($\alpha_1<0$ is
the flat or decreasing mass function we have already excluded above) then we
require $R_{0.5}>0.8$~pc in order not to evolve the mass ratio distribution of
solar type binaries too much. This, however, implies $f_{\rm tot}>f_{\rm
tot}^{\rm obs}$.

While firm constraints on $\alpha_1$ and $R_{0.5}$ are not possible
using these arguments because
the observational constraints are still too poor, our gedanken experiment does
suggest that there may exist a relationship between $R_{0.5}$ and $\alpha_1$,
i.e. between the stellar number density at birth and the initial mass
function. This line of thought implies that there may exist variation of the
stellar mass function in different dynamical birth configurations.
Podsiadlowski \& Price (1992) obtain a qualitatively similar result. They
suggest that the form of the mass function may be determined (at least in part,
given their simple physical model) by a competition between accretion and
proto-stellar collision rates. An insightful discussion of the physical factors
which might influence the spectrum of stellar masses formed can also be found
in Zinnecker, McCaughrean \& Wilking (1993).

Returning to our line of thought, three
distinct modes of star formation may have been identified (although it is
likely that they form a continuous sequence): very low star formation
efficiencies (the distributed mode), low star formation efficiencies (embedded
clusters, i.e. the alleged dominant mode, and Galactic clusters with somewhat
higher
efficiencies) and high efficiencies (Globular clusters). Here high efficiency
is equivalent to a deep potential well of the star-forming region so that only
a relatively small part of the original gas can be blown out.
The variations of the
densities within any one of these modes may be too small to imply significant
variations in the initial mass function. Thus the stellar initial mass function
may not be significantly different in Galactic and embedded clusters.
The differences in stellar
densities at birth between the three `distinct' modes may well be large enough
to provide
distinct initial mass functions implying that globular clusters may have
significantly different initial mass functions than Galactic or embedded
clusters, or a stellar population born in the distributed mode. Given these
speculations, we remember however
that Leinert et al. (1993) have found evidence that the mass ratio distribution
of pre-main sequence binaries born in the distributed mode is best approximated
with the Galactic field stellar mass function, which by its definition must be
the result from the dominant mode of star formation. This finding casts doubt
on the expected variation of the stellar mass function with star formation
mode, but as yet the data are too poor to reject it.

In the light of our findings it is illustrative to turn our argumentation
around.
If observations of star forming regions affirm that the majority of
stars form in embedded clusters that are dynamically equivalent to our dominant
mode cluster
then we have found an argument for the KTG(1.3) mass function
which is independent
of star count data and only rests on the correction of the mass-ratio
distribution of G~dwarf binaries for stimulated evolution in the
dominant mode cluster.

\nobreak\vskip 10pt\nobreak
\noindent{\bf 6.3 The initial distribution of periods}
\nobreak\vskip 10pt\nobreak
\noindent
Our initial period distribution was derived not by
matching to the observational pre-main sequence data but by inverting the
period depletion
function (Fig.~6). The derived initial period distribution, which is valid for
star formation in the dominant mode cluster ($R_{0.5}\approx0.8$~pc),
agrees with the pre-main sequence observational
constraints (Fig.~8) which have been obtained for distributed star formation
(assuming incompleteness corrections do not significantly alter the observed
distributions -- see Section~1). This result suggests that the initial period
distribution may be
representative of star formation in general, and may not be a strong function
of the star forming environment. This conclusion does not support the contrary
assertion by Durison \& Sterzik (1994), but future observations are needed to
clarify this issue.

\nobreak\vskip 10pt\nobreak
\noindent{\bf 6.4 The dominant mode embedded cluster}
\nobreak\vskip 10pt\nobreak
\noindent
The simulations of the dynamical
evolution of our aggregates assumes virial equilibrium without any additional
mass apart from that contributed by the stars. This is not a realistic model of
embedded clusters which loose most (70--90~per cent) of the natal material
through proto-stellar
outflows and winds, as discussed in greater length by Mathieu (1986).
Battinelli \& Capuzzo-Dolcetta (1991) find
that the majority of clusters have a lifetime of about 10~Myrs by which time
most of the natal gas must have been removed rendering the cluster unbound.
The orbital parameters of the young
binary population will thus be `frozen' after roughly $10^7$~yrs.

Lada \& Lada
(1991) define a cluster as a group of stars having a mass density of at least
$1\,M_\odot$~pc$^{-3}$. Following this definition the aggregates considered
here (Table~1) have a lifetime of about 250~Myrs (K3). Realistic
models of embedded clusters must therefore include a background potential which
evolves on a timescale of a few Myrs.
Lada, Margulis \& Dearborn
(1984), Pinto (1987) and Verschueren \& David (1989) consider the effects of a
changing background potential on
the stability of embedded clusters, but application to an initial aggregate of
binary systems awaits to be done.

By adding a background potential but keeping $R_{0.5}$ constant we increase the
velocity dispersion in the cluster (assuming virial equilibrium). Since the
cross section for collisions between binary systems is proportional to
$v^{-2}$, where $v$ is the relative velocity of two binaries well before the
encounter (see equation~26 in Heggie \& Aarseth 1992)
we expect stimulated evolution to decrease. Thus we need to decrease $R_{0.5}$
to increase the number density and thus the collision rates between the binary
stars.

The destruction rate of binary systems is

$${dn_{\rm b}\over dt}\propto -{n_{\rm b}^2\over v}, \eqno (12)$$

\noindent (equation~26 in Heggie \& Aarseth 1992), where
$n_{\rm b}\propto (N_{\rm bin}/R_{0.5}^3)$ is
the number density of binary stars. Assuming a star formation
efficiency $\epsilon$ we have a total birth aggregate mass $M_{\rm
tot}(0)=M_{\rm
stars}/\epsilon = N_{\rm bin}\,\overline{m}/\epsilon$, where $\overline{m}$ is
the
mean binary system mass ($0.64\,M_\odot$). The velocity dispersion in a
cluster is
$v\propto \left({M_{\rm tot}/R_{0.5}}\right)^{1\over2}$
so that

$$v_{\rm new}^2 = {1\over\epsilon}v_{\rm old}^2, \eqno (13)$$

\noindent where `new' refers to the dominant mode embedded cluster, and `old'
refers to the dominant mode cluster.

Assuming that the `dominant mode embedded cluster' has a lifetime
$\eta$ times smaller than our `dominant mode cluster' [$(N_{\rm
bin},R_{0.5})=(200,0.8\,{\rm pc})$] we need to `speed up' the destruction of
binary systems by a factor of $\eta$, i.e.
${dn_{\rm b,new}/dt} = \eta\,dn_{\rm b,old}/dt$.
We obtain
$n_{\rm b,new}^2 = (\eta/\epsilon^{1\over 2})\,n_{\rm b,old}^2$.
Thus

$${R_{0.5,{\rm new}}\over R_{0.5,{\rm old}}} =
{\epsilon^{1\over12}\over\eta^{1\over6}}
        \left({N_{\rm bin,new}\over N_{\rm bin,old}}\right)^{1\over3}. \eqno
(14)$$

\noindent If $N_{\rm bin,new}=N_{\rm bin,old}=200$ binaries form in an
aggregate and $R_{0.5,{\rm old}}=0.8\,$~pc (the dominant mode cluster) then we
obtain Table~3 for $R_{0.5,{\rm new}}$:

\bigbreak
\vskip 3mm
\bigbreak

\hang{ {\bf Table 3.} Expected half mass radii $R_{0.5,{\rm new}}$ of the
dominant mode embedded cluster

\nobreak
\vskip 1mm
\nobreak
{\hsize 11 cm \settabs 11 \columns

\+&&$\eta$ &&5 &&10 &&20 &&30\cr
\+&&$\epsilon$\cr
\+&&0.1 &&0.505~pc &&0.450~pc &&0.401~pc &&0.375~pc\cr
\+&&0.2 &&0.534    &&0.477    &&0.425    &&0.397   \cr
\+&&0.3 &&0.553    &&0.493    &&0.439    &&0.411\cr
\+&&0.4 &&0.567    &&0.505    &&0.450    &&0.421\cr

}

\bigbreak
\vskip 3mm
\bigbreak

\noindent Thus $0.37$~pc$<R_{0.5,{\rm new}}<0.57$~pc {\it if} we demand
{\it dynamical equivalence} with our dominant mode cluster!
These $R_{0.5,{\rm new}}$ are `initial' values,
i.e. {\it at the time of star--gas decoupling} (see also Section~6.6),
and the observed embedded
clusters are likely to have already expanded from their
initial configuration. If initially $R_{0.5}=0.37$~pc then the
approximate average inter-binary spacing (approximately 0.1~pc) corresponds to
a period of $10^{9.14}$~days for a binary
system of total mass $0.64\,M_\odot$. This is larger than the maximum
period in our initial period distribution, and so {\it we do not expect erosion
of the period distribution alone from crowding} in the dominant mode embedded
cluster.

The results of our simulations which neglect the gas component thus
appear to be robust
because $R_{0.5,{\rm new}}$ does not change significantly even if the effect of
the
(initially mass dominating) gas component is taken into account.
$R_{0.5,{\rm new}}$ is not significantly different to $R_{0.5,{\rm
old}}=0.8$~pc. Numerical simulations will, however, be necessary to verify
these conclusions.

\nobreak\vskip 10pt\nobreak
\noindent{\bf 6.5 Small groups of stars as the dominant mode of star
formation?}
\nobreak\vskip 10pt\nobreak
\noindent
In Section~6.2 we have discussed some alternatives to our model (e.g.: a
different mass
function and distributed star formation being the most common mode) which we
find are not consistent with the observational data. In this context
we discuss the possibility that binaries may form preferentially in small
groups. McDonald
\& Clarke (1993) assume stars form in
small goups of three to ten and that the most massive stars in these groups
form binary systems by dynamical capture.
They obtain a mass ratio
distribution as observed by Duquennoy \& Mayor (1991) despite using a mass
function similar to that adopted here (equation~1),
as well as a decreasing proportion of binary systems with decreasing primary
mass. The overall proportion of binaries thus formed is, however, much smaller
than $f_{\rm tot}^{\rm obs}\approx0.5$.

Energy dissipation in the
remaining circum-protostellar material is likely to be very important in
raising the proportion of binaries formed by capture. McDonald \& Clarke (1995)
include a simple model of energy dissipation in
circumstellar discs. They obtain a larger proportion of binaries
than in the absence of discs (McDonald \& Clarke 1993). The proportion of
G- and M-dwarf binaries of the resulting stellar system population is about
correct if the preferred initial group size is $N\approx10$ (their fig.~13).
If stars are distributed randomly throughout the group then the component
masses in binary systems are uncorrelated (the probability of two closest
neighbours dissipating enough relative kinetic energy to form a bound binary is
increased). This is shown in their fig.~11 but is not consistent with the
main sequence G~dwarf mass ratio distribution (Kroupa \& Tout 1992).
This result would appear to indicate that dissipative capture in small groups
($N\approx10$) of proto stars may not be able to account for the distribution
of dynamical properties of stellar systems in the Galactic disk.

In Section~6.6 we speculate that typical star formation in the Galactic disk
may,
however, be characterised by the formation of aggregates containing tens of
dissipative small-$N$ groups.

\nobreak\vskip 10pt\nobreak
\noindent{\bf 6.6 The star formation process}
\nobreak\vskip 10pt\nobreak
\noindent
For star formation (i.e. for times {\it during} the first $10^5-10^6$~yrs) we
may interpret our results as follows:

Our assumptions and results would appear to be valid if most proto-stellar
binary systems would form by dissipative capture in small groups of about
ten accreting proto-stars that are clustered in aggregates of tens of such
groups (Section~6.5). Aarseth \& Hills (1972) show that
subclustering enhances the proportion
of binary systems formed by capture even if stars are approximated as point
particles. Murray \& Clarke (1993, in their section~5) similarly find a higher
production of binary systems by dissipative capture in clumpy initial
conditions, the virial velocities in each subgroup being sufficiently small to
imply capture rather than disc destruction.
This may produce uncorrelated component masses (see Section~6.5),
a dynamically
relaxed eccentricity distribution and
a majority of proto-binaries with large periods. If the evolution timescale of
the individual groups is sufficiently smaller than the evolution timescale of
the whole aggregate (as is the case in the simulations of Aarseth \& Hills
1972) then
we would arrive at our results. For example, a group with $R_{0.5}=0.04$~pc
consisting of 10 stars has a relaxation time (Spitzer \& Hart 1971)
of about $T_{\rm relax}=0.07$~Myrs,
whereas an aggregate with $R_{0.5}=0.8$~pc consisting of 200 binaries has
$T_{\rm relax}=10$~Myrs.

Groups of about 10~accreting protostars may result from fragmentation
during protostellar collapse
and fragmentation of accretion disks.
We need magnetohydrodynamics to study the very early evolution, and in fact the
onset of star formation itself (see e.g. Patel \& Pudritz 1994).
Modern computational facilities, and in part our knowledge of the relevant
physics (e.g. opacities, radiation transfer etc.) do not presently allow this
time scale to be studied in great detail. However, despite the present
limitations
the Cardiff group has been obtaining interesting numerical results on the
formation
of binary systems and stellar aggregates in shocked and colliding molecular
cloud
clumps (see e.g. Turner et al. 1994).
It remains unclear whether the proto-stellar clumps remain in
virial equilibrium in their cloud core, or whether they freeze out of the core
with smaller velocities than the characteristic velocity dispersion associated
with the
turbulent molecular gas, whether a violent cold collapse occurs, or whether the
proto-stars sink to the centre of the potential well with dynamical friction
(Just, Kegel \& Deiss 1986)
reducing their rate of infall, and how the gaseous medium reacts,
i.e. how
much energy is deposited from the motion of the proto-stars into the molecular
gas, which in turn affects the star--gas coupling and gas removal.

The protostars will decouple from the gas
after the first $10^5-10^6$~yrs have passed.
This occurs because the gas density remaining
within the young aggregate decreases owing to gas accretion onto the stars and
gas removal by very young active pre-main sequence stars.

The results of our simulations uncover the most common dynamical structures
which may be established {\it after} this early time.

\nobreak\vskip 10pt\nobreak
\noindent{\bf 6.7 Hierarchical star formation?}
\nobreak\vskip 10pt\nobreak
\noindent
Our speculation in Section~6.6 suggests that the notion that aggregates of
groups of about 10 proto stars may condense out of a molecular cloud core may
be  consistent with our assumptions (Section~3.2.1).

It is now known that in the Taurus--Auriga star forming regions pre-main
sequence stars are distributed spatially in a fractal or self-similar way (see
e.g. Larson 1995). The stars are clustered hierarchically, with smaller
groupings within larger ones over a considerable range of scales.
This self-similar clustering breaks down on scales smaller than
0.04~pc, most if not all stars being members of binary systems.
Taurus--Auriga
is a low-density star forming environment, i.e. the stars have formed in the
distributed or isolated mode.

Our inverse dynamical population synthesis suggests
that most stars in the Galactic
disk may have formed in embedded clusters rather than in the distributed mode.
The properties of the binary star population in embedded clusters are unknown.
Inverse dynamical population synthesis suggests that after star--gas
decoupling most stars may be
in about 1~pc large aggregates of a few hundred binary systems.

Binary
systems would thus appear to be the primary subunit of star formation in {\it
both} the isolated {\it and} clustered star formation mode.

It is thus possible
that star formation produces hierachical clustering which
evolves, after a few dynamical timescales of the various fractal subunits, to
the
structures that about 1--10~Myr old populations are observed to have.
Low-density star forming environments may evolve
significantly only on the smallest scales thereby producing a large proportion
of binaries. The high-density star forming regions (the `clustered mode') would
contract to embedded clusters.
If this is correct then the initially hierarchical
structure would first produce predominantly binary systems with uncorrelated
component masses, a dynamically relaxed eccentricity distribution and
predominantly long periods. These binary systems would then interact in the
evolving aggregate to produce the Galactic disk stellar population after the
aggregate has disintegrated after approximately $10^7$~yrs.
In even higher density environments with relatively high star formation
efficiency the initial hierarchical distribution may evolve to Galactic
clusters.

\bigskip
\bigbreak
\noindent{\bf 7 CONCLUSIONS}
\nobreak\vskip 10pt\nobreak
\noindent
Recent observational data (Figs.~1 and~2) suggest that the distribution of
orbital periods of pre-main-sequence binary systems differ from main sequence
orbital distributions such that star-forming regions have a significantly
higher binary proportion.

We assume that all stars form in aggregates of~200 binary systems (a
reasonable first assumption, see Section~3.2.2) with component
masses paired at random from the KTG(1.3) mass function derived from
a thorough analysis of local and deep star count data by Kroupa et al.
(1993). This mass function contradicts the observed mass function of
secondaries in binaries with G~dwarf primaries which has a
deficit of low mass companions (Kroupa \& Tout
1992), but it is consistent with the distribution of brightness ratios for
pre-main sequence binaries (Leinert et al. 1993). These assumptions will be
verified by observations if our results are a reasonable approximation of the
star-formation process.

Assuming an initially
flat log$_{10}P$ distribution ($P$ in days, $3\le $log$_{10}P\le 7.5$), which
is
consistent with the observed pre-main sequence period distribution, we find
that a
cluster with half mass radius $0.25\,{\rm pc}<R_{0.5}<0.8\,$pc has the correct
dynamics to lead to a proportion of main sequence binary systems
consistent with the observed value
($47\pm5$~per~cent, Fig.~3). A cluster with $R_{0.5}=0.8\,$pc
leads to a mass-ratio distribution for binaries with G-dwarf primaries as
observed (Fig.~4), and to a
depletion of the period distribution with increasing period which
is similar to the observed main sequence distribution (Fig.~5).

This finding suggests that the dominant mode of star
formation in the Galactic disk may be clustered star formation. We refer to our
$R_{0.5}\approx0.8$~pc cluster as the {\it dominant mode cluster}. We estimate
that the initially mass dominating gas component in embedded clusters does not
significantly affect this result (Section~6.4).
Presently available observations of star forming regions appear to confirm our
result (Section~6.1), but the sample of star-forming regions has to be
extended before this result can be taken as conclusive.

Hardening and softening of binary systems in the stellar aggregates considered
here (Table~1) does not significantly change the numbers of orbits with
log$_{10}P<3$ and log$_{10}P>7.5$, respectively (Section~4).
The orbital depletion
function can be inverted to derive an initial distribution of periods
(Section~5). We term
the evolution of orbital elements due to mutual perturbation of neighbouring
systems {\it stimulated evolution}.

The observational data of pre-main sequence binaries are valid only for
distributed
or isolated star formation. We find that the initial period distribution
(equation~11) is very
similar to the observed period distribution of pre-main sequence binaries.
This initial period distribution was derived for clustered star formation.
Significant variation of binary star properties with star formation environment
is thus not suggested by the results obtained here but future investigation is
required to confirm this assertion. If it is found that the observed orbital
distributions in different
star forming regions differ then we caution against inferring that the
primordial binary star properties depend on local conditions (Durison \&
Sterzik 1994) unless it can be
established that the observed difference cannot be due to stimulated evolution.
In comparing stellar dynamical properties (mass function, orbital parameters)
in different star-forming regions it is important to establish the stage of
dynamical evolution of these regions. For example, in K3 we argue that a
distributed population of young stars in the L1641 molecular cloud may have
originated in by now dissolved aggregates. If this is true then this population
will have a binary
proportion of 50--60~per cent and a period distribution similar to the main
sequence distribution. These properties (if verified) would be different to the
Taurus--Auriga pre-main sequence binaries.

We find evidence that the mass function power law index $\alpha_1$ (equation~1)
may depend on the stellar number density at birth (Section~6.2). Our
assumptions and results appear to be consistent with hierarchically distributed
star formation (Section~6.7).

We conclude with the following  question pertaining to the origin of stars in
the Galactic disk: Once star formation begins, have most of the molecular cloud
cores evolved in parameter space (pressure, temperature, turbulent
kinetic energy etc.) in such a way
that the subsequent dynamics of the aggregate of stars that freeze out of the
core is
{\it equivalent} to that of our dominant mode cluster [$(N_{\rm
bin},R_{0.5})\approx(200,0.8\,{\rm pc})$] (Section~4.3)?
It would appear to be of great interest to conduct additional simulations with
the period, eccentricity and mass ratio distributions of binaries at birth
derived here but other virial ratio (e.g. violent relaxation), changing
background potential leading
to shorter time scales for stimulated evolution, and other mass and/or size of
the initial aggregate of binaries, to seek those initial
configurations which give the same final distributions of stellar dynamical
properties as are observed
for Galactic field stars and which are obtained with our dominant mode cluster.
An initial non-negligible proportion of triple and quadruple systems will be
required to explain the number of such systems in the Galactic disk and in star
forming regions (K2).

This may be extended to include non-dominant modes at the present epoch of
star formation, such
as rich Galactic clusters or globular clusters, to investigate if our initial
stellar dynamical properties are consistent with the
observed stellar dynamical properties in such clusters.

\bigskip
\bigbreak
\par\noindent{\bf ACKNOWLEDGMENTS}
\nobreak
\nobreak
\noindent I am very grateful to Sverre Aarseth without whose
kind
help with and distribution of {\it nbody5} this work would not have been
possible.
The flux of subroutines from Cambridge to
Heidelberg was partially, but probably not sufficiently, countered by a flow
of wine in the opposite direction.
I am also grateful to Cathy Clarke for very useful discussions and for
pointing out to me the research of Lada et al. after a first draft of this
paper, and to Peter Schwekendiek for his rapid acquaintance with and
maintenance of the new computing facilities installed at ARI towards the end of
1993.

\bigskip
\noindent{\bf REFERENCES}
\nobreak
\bigskip
\nex Aarseth, S. J., Hills, J. G., 1972, A\&A 21, 255
\nex Aarseth, S. J., Henon, M., Wielen, R., 1974, A\&A, 37, 183
\nex Aarseth, S. J., 1992, N-Body Simulations of Primordial Binaries and Tidal
     Capture in Open Clusters. In:
     Duquennoy, A., Mayor, M. (eds), Binaries as Tracers of Stellar
     Formation, Cambridge Univ. Press, Cambridge, p.6
\nex Aarseth, S. J., 1994, Direct Methods for N-Body Simulations. In:
      Contopoulos, G., Spyrou, N. K., Vlahos, L. (eds.), Galactic Dynamics and
      N-Body Simulations, Springer, Berlin, p.277
\nex Abt, H. A., Levy, S. G., 1976, ApJS 30, 273
\nex Abt, H. A., 1987, ApJ 317, 353
\nex Battinelli, P., Capuzzo-Dolcetta, R., 1991, MNRAS 249, 76
\nex Battinelli, P., Brandimarti, A., Capuzzo-Dolcetta, R., 1994, A\&AS 104,
     379
\nex Bodenheimer, P., Ruzmaikina, T., Mathieu, R. D., 1993, Stellar Multiple
     Systems: Constraints on the Mechanisms of Origin, In: Levy, E. H., Lunine,
     J. I., Matthews, M. S. (eds), Protostars and Planets III, Univ. of Arizona
     Press, Tucson, p.367
\nex Bonnell, I., 1994, Fragmentation and the Formation of Binary and Multiple
     Systems, In: Clemens, D., Barvainis, R. (eds), Clouds, Cores, and Low Mass
     Stars, ASP Conf. Series, in press
\nex Bonnell, I., Bastien, P. 1992, The Formation of Non-Equal Mass Binary and
     Multiple Systems. In:
     McAlister, H. A., Hartkopf, W. I. (eds.) Complementary Approaches to
     Double and Multiple Star Research, Proc. IAU Coll. 135, ASP Conference
     Series, 32, p.206
\nex Boss, A., 1988, Comments Astrophys. 12, 169
\nex Burkert, A., Bodenheimer, P., 1993, MNRAS 264, 798
\nex Carpenter, J. M., Snell, R. L., Schloerb, F. P., Strutskie, M. F., 1993,
     ApJ 407, 657
\nex Close, L. M., Richer, H. B., Crabtree, D. R., 1990, AJ 100, 1968
\nex Duquennoy, A., Mayor, M., 1991, A\&A 248, 485
\nex Durisen, R. H., Sterzik, M. F., 1994, A\&A 286, 84
\nex Elmegreen, B. G., 1993, ApJ 419, L29
\nex Fischer, D. A., Marcy, G. W., 1992, ApJ 396, 178
\nex Ghez, A. M., Neugebauer, G., Matthews, K., 1993, AJ 106, 2005
\nex Goldberg, D., Mazeh, T., 1994, A\&A 282, 801
\nex Heggie, D. C., 1975, MNRAS 173, 729
\nex Heggie, D. C., Aarseth, S. J., 1992, MNRAS 257, 513
\nex Hills, J. G., 1975, AJ 80, 809
\nex Just, A., Kegel, W. H., Deiss, B. M., 1986, A\&A 164, 337
\nex Kroupa, P., 1995a, The Dynamical Properties of Stellar Systems in the
     Galactic Disc, MNRAS, in press (K2)
\nex Kroupa, P., 1995b, Star Cluster Evolution, Dynamical Age Estimation and
     the Kinematical Signature of
     Star Formation, MNRAS, in press (K3)
\nex Kroupa, P., 1995c, Unification of the Nearby and Photometric Stellar
     Luminosity Functions, ApJ 453, 0000
\nex Kroupa, P., Tout, C. A., 1992 MNRAS, 259, 223
\nex Kroupa, P. Tout, C. A., Gilmore, G., 1990, MNRAS 244, 76
\nex Kroupa, P., Tout, C. A., Gilmore, G., 1993, MNRAS 262, 545
\nex Lada, C. J., Margulis, M., Dearborn, D., 1984, ApJ 285, 141
\nex Lada, C. J., Lada, E. A., 1991, The Nature, Origin and Evolution of
     Embedded Star Clusters. In: James, K. (ed.), The Formation and Evolution
     of Star Clusters, ASP Conf. Series 13, 3
\nex Larson, R. B., 1995, MNRAS 272, 213
\nex Leinert, Ch., Zinnecker, H., Weitzel, N., Christou, J., Ridgway, S. T.,
     Jameson, R., Haas, M., Lenzen, R., 1993, A\&A 278, 129
\nex Mathieu, R. D., 1986, Highlights of Astronomy, 7, 481
\nex Mathieu, R. D., 1992, The Short-Period Binary Frequency Among Low-Mass
     Pre-Main Sequence Stars. In:
     McAlister, H. A., Hartkopf, W. I. (eds.) Complementary Approaches to
     Double and Multiple Star Research, Proc. IAU Coll. 135, ASP Conference
     Series, 32, p.30
\nex Mathieu, R. D., 1994, ARA\&A 32, 465
\nex Mayor, M, Duquennoy, A., Halbwachs, J.-L., Mermilliod, J.-C.,
     1992, CORAVEL Surveys to Study Binaries of Different Masses and Ages. In:
     McAlister, H. A., Hartkopf, W. I. (eds.) Complementary Approaches to
     Double and Multiple Star Research, Proc. IAU Coll. 135, ASP Conference
     Series, 32, p.73
\nex Mazeh, T., Goldberg, D., 1992, ApJ 394, 592
\nex Mazeh, T., Goldberg, D., Duquennoy, A., Mayor, M., 1992, ApJ 401, 265
\nex McCaughrean, M. J., Stauffer, J. R., 1994, AJ 108, 1382
\nex McDonald, J. M., Clarke, C. J., 1993, MNRAS 262, 800
\nex McDonald, J. M., Clarke, C. J., 1995, The Effect of Star-Disc Interactions
     on
     the Binary Mass-Ratio Distribution, MNRAS, in press
\nex McMillan, S., Hut, P., 1994, ApJ, 427, 793
\nex Mikkola, S., Aarseth, S. J., 1990, Celest. Mech. Dyn.
     Astron. 47, 375
\nex Mikkola, S., Aarseth, S. J., 1993, Celest. Mech. Dyn.
     Astron. 57, 439
\nex Murray, S. D., Clarke, C. J., 1993, MNRAS 265, 169
\nex Patel, K. \& Pudritz, R. E., 1994, ApJ 424, 688
\nex Pinto, F., 1987, PASP 99, 1161
\nex Price, N. M., Podsiadlowski, P., 1995, MNRAS 273, 1041
\nex Pringle, J. E., 1989, MNRAS 239, 361
\nex Reipurth, B., Zinnecker, H., 1993, A\&A 278, 81
\nex Richichi, A., Leinert, Ch., Jameson, R., Zinnecker, H., 1994, A\&A 287,
     145
\nex Simon, M., 1992, Multiplicity Among the Young Stars. In:
     McAlister, H. A., Hartkopf, W. I. (eds.) Complementary Approaches to
     Double and Multiple Star Research, Proc. IAU Coll. 135, ASP Conference
     Series, 32, p.41
\nex Spitzer, L, Hart, M. H., 1971, ApJ 164, 399
\nex Strom, K. M., Strom, S. E., Merrill, K. M., 1993, ApJ 412, 233
\nex Tout, C. A., 1991, MNRAS 250, 701
\nex Turner, J. A., Bhattel, A. S., Chapman, S. J., Disney, M. J.,
     Pongracic, H., Whitworth, A. P., 1994, The Formation of a
     ``Protocluster''. In: Franco, J.,
     Lizano, S., Aguilar, L., Daltabuit, E. (eds.), Numerical Simulations in
     Astrophysics, Cambridge Univ. Press, Cambridge, p.345
\nex Verschueren, W., David, M., 1989, A\&A 219, 105
\nex Zinnecker, H., McCaughrean, M. J., Wilking, B. A., 1993, The Initial
     Stellar Population. In:
     Levy, E. H., Lunine,
     J. I., Matthews, M. S. (eds), Protostars and Planets III, Univ. of
     Arizona Press, Tucson,
     p. 429

\vfill\eject

\centerline{\bf Figure captions}
\smallskip
\noindent
{\bf Figure 1.} Collation of all currently available data on the period
distribution of late-type binary systems (Section~2). Solid dots show the
distribution
of orbits for solar-mass main sequence binary systems (Dyquennoy \& Mayor
1991), open circles indicate the preliminary distribution of periods
for K-dwarf main sequence binary systems (Mayor et al. 1992) and stars
represent M~dwarf binaries (Fischer \& Marcy 1992). The area under each
distribution is the proportion of orbits among all~G, K and~M stellar systems.
The distribution
of orbital periods of low-mass pre-main sequence stars, obtained by
transforming from the projected separations, are shown by open triangles (Simon
1992), open squares (Leinert et al. 1993 and Richichi et al. 1994), open
star (Reipurth
\& Zinnecker (1993) (on top of an open square at log$_{10}P=6.5$) and cross
(Ghez et al. 1993). The proportion of pre-main sequence
binaries with log$_{10}P<2$ is indistinguishable from the Galactic field
distribution (Mathieu 1992, 1994).

\vskip 5mm

\noindent {\bf Figure 2.} {\bf Top panel}: the distribution of mass ratios
for main
sequence solar-type systems of all periods (Section~2). The filled dots
show the
distribution derived by Duquennoy \& Mayor (1991).
The open circles denote their mass ratio distribution
corrected here for the bias in short-period binaries (Mazeh \& Goldberg 1992)
by adding the corrected short- and long-period distributions shown in the
middle panel. The
upper dotted, middle dashed and lower dotted curves represent the
KTG($\alpha_1$) mass
function with $\alpha_1=1.85,1.3,0.7$ (equation~1), respectively, which
is the 95~per~cent confidence interval for $\alpha_1$ (Kroupa et al.
1993). The units of the ordinate are the number of systems per bin.
{\bf Middle panel}: the corrected distribution for short period
(log$_{10}P<3.5$) binaries constructed from the data given by Mazeh et al.
(1992) is depicted by open circles. The one-sigma error range is
indicated by the dotted lines. The distribution of orbits for long periods
(log$_{10}P>3.5$) is shown by the filled dots and was obtained from the data
presented by Duquennoy \& Mayor (1991). Both distributions are normalised to
unit area. {\bf Bottom panel}:
the short period (log$_{10}P<3$, open circles) and long period
(log$_{10}P>3$, solid circles) eccentricity distributions from Duquennoy \&
Mayor (1991), both normalised to unit area.

\vskip 5mm

\noindent {\bf Figure 3.} Evolution of the total binary fraction with time for
the first four binary star aggregates in Table~1 (Section~4.2). Binary stars
also
form in the aggregates by capture, and the proportion of such binaries is shown
for the single star aggregates (Table~1 with $N_{\rm bin}=0$).

\vskip 5mm

\noindent {\bf Figure 4.}
The initial (dotted histogram) and final (solid
histogram) distribution of secondary
masses for primaries with a mass in the range
$0.85\,M_\odot<m_1<1.1\,M_\odot$ for
the first four binary star aggregates in Table~1
(Section~4.2).  The observational data
in the top panel of Fig.~2 have been scaled to the initial models at
$m_2>0.6\,M_\odot$.

\vskip 5mm

\noindent {\bf Figure 5.} The initial (dotted histogram) and final
(solid histogram) distribution of periods for
the first four binary star aggregates in Table~1
(Section~4.2).  The observational data are as in Fig.~1.

\vskip 5mm

\noindent {\bf Figure 6.} The depletion function $c_{\rm P,i}(1\,{\rm Gyr})$
(equation~8) introduced in Section~5.1 for the first four aggregates
(Table~1): open stars ($R_{0.5}=2.53$~pc), solid circles ($R_{0.5}=0.77$~pc),
open circles ($R_{0.5}=0.25$~pc) and crosses ($R_{0.5}=0.08$~pc). $c_{\rm
P,i}>1$ implies a gain in the ith log$_{10}P$ bin.

\vskip 5mm

\noindent {\bf Figure 7.} The corrected inital period distributions
(equation~9a) obtained
in Section~5.1 by applying the depletion function of Fig.~6 to the
main
sequence data of Duquennoy \& Mayor (1991) shown here as solid dots. Results
for each of the first four aggregates (Table~1) are shown:
dotted line
($R_{0.5}=0.08\,$pc), long dashed line ($R_{0.5}=0.25\,$pc), solid line with
large solid dots ($R_{0.5}=0.77\,$pc) and dot-dashed line ($R_{0.5}=2.53\,$pc).
The thick solid line is the initial period distribution adopted as the
first iteration (equation~10).

\vskip 5mm

\noindent {\bf Figure 8.} a) First and b) second iterations towards the
initial period distribution. In both panels the continuous initial
distributions (equations~10a and~11a) are shown as the thick solid line. The
distribution of orbits picked initially from these distributions are shown by
the dotted
histograms, and the final period distributions are shown by
the solid histogram. The observational data are as in Fig.~1.

\vfill
\bye